\documentclass[useAMS,usenatbib]{mn2e}
\usepackage{txfonts,graphicx}
\usepackage{hyperref}
\usepackage{hyphenat}
\hypersetup{ 
	bookmarks=true,         
    colorlinks=true,        
  linkcolor=blue,          
      citecolor=blue,        
      urlcolor=blue,           
}
\usepackage[switch]{lineno}

 \title{MAGIC detection of short-term variability of the high-peaked BL Lac object 1ES 0806+524}
\author[J.~Aleksi\'c~et.~al.]
{
\parbox{\textwidth}{
\small{
J.~Aleksi\'c$^{1}$,
S.~Ansoldi$^{2}$,
L.~A.~Antonelli$^{3}$,
P.~Antoranz$^{4}$,
A.~Babic$^{5}$,
P.~Bangale$^{6}$,
J.~A.~Barrio$^{7}$,
J.~Becerra Gonz\'alez$^{8,\,25}$,
W.~Bednarek$^{9}$,
E.~Bernardini$^{10}$
% ...
}}
 \vspace{0.25cm}\\
(Affiliations can be found at the end of the article).}
\begin{document}
\hyphenation{VERITAS}
\date{Accepted 2015 April 20. Received 2015 April 13; in original form 2014 November 12}
\pagerange{\pageref{firstpage}--\pageref{lastpage}}\pubyear{2015}
\maketitle
\label{firstpage}
 \begin{abstract}
 \small

The high-frequency-peaked BL Lac (HBL) 1ES 0806+524 (z = 0.138) was discovered in VHE $\gamma$ rays in 2008. Until now, the broad-band spectrum of 1ES 0806+524 has been only poorly characterized, in particular at high energies. We analysed multiwavelength observations from $\gamma$ rays to radio performed from 2011 January to March, which were triggered by the high activity detected at optical frequencies. These observations constitute the most precise determination of the broad-band emission of 1ES 0806+524 to date. The stereoscopic MAGIC observations yielded a $\gamma$-ray signal above 250\,GeV of $(3.7 \pm 0.7)$ per cent of the Crab Nebula flux with a statistical significance of 9.9\,$\sigma$. The multiwavelength observations showed significant variability in essentially all energy bands, including a VHE $\gamma$-ray flare that lasted less than one night, which provided unprecedented evidence for short-term variability in 1ES 0806+524. The spectrum of this flare is well described by a power law with a photon index of $2.97 \pm 0.29$ between $\sim$150\,GeV and 1\,TeV and an integral flux of $(9.3 \pm 1.9)$ per cent of the Crab Nebula flux above 250\,GeV. The spectrum during the non-flaring VHE activity is compatible with the only available VHE observation performed in 2008 with VERITAS when the source was in a low optical state. The broad-band spectral energy distribution can be described with a one-zone Synchrotron Self Compton model with parameters typical for HBLs, indicating that 1ES 0806+524 is not substantially different from the HBLs previously detected.
\end{abstract}

\begin{keywords}
BL Lacertae objects: individual: 1ES 0806+524 - galaxies: active - galaxies: jets - gamma rays: galaxies - radiation mechanisms: non-thermal
\end{keywords}

\section{Introduction}
Active galactic nuclei (AGN) are among the most variable objects in the known Universe. Their broad-band energy spectrum spanning from radio to very high energy (VHE, E $>$ 100\,GeV) $\gamma$ rays can be characterised by two distinct peaks: one in the sub-mm to X-ray range, commonly interpreted as synchrotron radiation, and a second one in the $\gamma$-ray band that is hypothesised to originate from inverse Compton scattering of photons. The emission is assumed to originate from relativistic particle jets launched along the axis of the accretion disc of matter that surrounds a supermassive black hole (e.g.~\citealt{blandford74}). Numerous AGNs are known to be blazars, characterised by relativistic jets closely aligned to the line of sight of the observer. Based on their optical spectra, blazars are divided into two classes: flat spectrum radio quasars (FSRQ) that show broad emission lines, and BL Lacertae objects (BL Lacs) characterised by the weakness or even absence of such emission lines~\citep{morris91,stickel91}. Depending on the frequency of the low-energy peak of the spectral energy distribution (SED) the latter class is subdivided into high- (HBL), intermediate- (IBL), and low- (LBL) frequency-peaking BL Lac objects~\citep{padovani95}. The Tuorla Blazar Monitoring Program~\citep{takalo08} monitors a sample of objects in the optical band, triggering VHE observations by the MAGIC (Major Atmospheric Gamma-Ray Imaging Cherenkov) telescopes during high optical states~\citep{albert06, albert07a, albert08a, anderhub09, aleksic12a, aleksic12b}. This allowed the discovery in VHE $\gamma$ rays of several blazars, especially HBLs and leads to the suggestion that in HBLs a high optical state is typically accompanied by a high state in VHE $\gamma$ rays. In some cases, studies of broad-band emission confirm a possible connection between the two wavebands (e.g.~\citealt{aharonian09,aleksic12c}) whereas in other sources this seems to be only partially the case (e.g.~\citealt{foschini07,foschini08}).

1ES 0806+524 (RA 08:09:49.18673, DEC 52:18:58.2507; J2000,~\citealt{petrov11}) is classified as a BL Lac~\citep{schachter} with a redshift of 0.138~\citep{bade98}. It was suggested as a VHE candidate ~\citep{costamante02} with a predicted intrinsic flux of F$_{E\mathrm{>0.3\,TeV}} = 1.36\times10^{-11}\,\mathrm{cm}^{-2}\,\mathrm{s}^{-1}$. Several VHE observations have been carried out by the Whipple Collaboration and the HEGRA Collaboration yielding flux upper limits above 300\,GeV~\citep{horan04,delacalle03} and 1.09\,TeV~\citep{aharonian04}, respectively.

In 2008, the VERITAS Collaboration reported the first detection of 1ES 0806+524 in the VHE $\gamma$-ray band~\citep{acciari09}. The collected data set spanned from 2006 November to 2008 April yielding a total of 245 excess events with a significance of 6.3\,$\sigma$. The integral flux above 300\,GeV corresponded to 1.8 per cent of the Crab Nebula flux, which is below earlier upper limits obtained by MAGIC-I observations in 2005 (5.6 per cent C.U.\footnote{The integrated flux level in Crab units (C.U.) is obtained by normalizing the integrated flux measured above a certain threshold to the Crab Nebula flux, which is considered to be stable, measured above the same threshold.} above 230\,GeV;~\citealt{albert08b}; 7.2 per cent C.U. above 140\,GeV;~\citealt{aleksic11}). No significant variability on a monthly timescale (the $\chi^2$/d.o.f.\footnote{d.o.f.= Degrees of freedom.} for a fit with a constant is 6.78/5;~\citealt{acciari09}) could be established for this object. The spectrum was obtained only for a subset of data taken with four telescopes during winter 2007/2008. Accordingly, the spectral characteristics were poorly defined. Thus, the spectral index was measured to be 3.6 $\pm$ 1.0$_{\mathrm{stat}}\pm$0.3$_{\mathrm{sys}}$ in a narrow range between 300\,GeV and 700\,GeV. The time-independent version of the one-zone jet radiation transfer code of~\citet{boettcher02}, with parameters appropriate for a pure Synchrotron Self Compton (SSC) model~\citep{maraschi03} was able to describe the data sufficiently well within the uncertainties.

In this paper we report the highly significant detection of this source with the MAGIC stereoscopic system during a flaring state in 2011 February. We analyse in detail the variability and the spectral evolution of the source and present complementary data from observations in high-energy (HE, 30\,MeV $<$ E $<$ 100\,GeV) $\gamma$-rays carried out by the \textit{Fermi} Large Area Telescope (LAT), in X-rays performed by the \textit{Swift} satellite and in the optical R-band by the Kungliga Vetenskapsakademien (KVA) telescope. Radio data coverage is provided by the Owens Valley Radio Observatory (OVRO) telescope at 15\,GHz. Thanks to the good multiwavelength (MWL) coverage, we investigate the connection among these wavebands. Finally, we model the SED including all MWL data.

\section{Observation and data analysis}

\subsection{MAGIC observations and data analysis}

Since 2009 MAGIC has operated as a stereoscopic system of two 17\,m Imaging Atmospheric Cherenkov Telescopes (IACTs) located at the Roque de Los Muchachos, La Palma, Canary Islands (28.8$^\circ$ N, 17.8$^\circ$ W, 2225\,m a.s.l.). Due to its low energy threshold (as low as 60\,GeV in normal trigger mode) and high sensitivity\footnote{Better than 0.8 per cent of the Crab Nebula flux in 50\,h of observing time above 290\,GeV~\citep{aleksic12d}.}, it is a well suited instrument for VHE $\gamma$-ray observations of blazars.

MAGIC observed 1ES 0806+524 between 2011 January and March on 13 nights for a total exposure of $\sim$24\,h. The observations from February to March were triggered by a flux increase in the optical R-band as part of a dedicated MAGIC Target of Opportunity (ToO) program to observe blazars showing high activity at other wavelengths. After applying quality selection cuts based on the event rate, $\sim$3.8\,h of data were discarded. Additionally, corrections for the dead time of the readout system yielded an effective observation time of 16.1\,h. Part of the observations were carried out under moderate moonlight conditions.

Observations were performed in the so-called \emph{wobble} mode~\citep{fomin94} during which both telescopes alternated every 20 minutes between two sky positions with an offset of 0.4$^\circ$ from the source. The acquired data cover a zenith angle from 23$^\circ$ to 46$^\circ$.

The data analysis was performed using the MAGIC standard tool `MARS'~\citep{moralejo09} including adaptations to stereoscopic observations. Based on the timing information~\citep{aliu09}, and absolute cleaning levels, the image cleaning was performed. The images were parametrized in each telescope individually according to the description of Hillas~\citep{hillas85}.

For the reconstruction of the shower arrival direction the random forest regression method (RF DISP method;~\citealt{aleksic10}) with the implementation of stereoscopic parameters such as the impact distance of the shower on the ground and the height of the shower maximum was used~\citep{lombardi11}.

To perform the gamma-hadron separation, the random forest method was applied~\citep{albert08c} using the image parameters of both telescopes and the shower impact point and height maximum. Energy lookup tables were used for the energy reconstruction. An angular resolution of $\sim$0.07$^\circ$ at 300 GeV and an energy resolution as good as 16 per cent in the medium energy range (few hundred GeV) are achieved (details on the stereo MAGIC analysis can be found in~\citealt{aleksic12d}).

For sources with VHE $\gamma$-ray spectra similar to that of the Crab Nebula, the sensitivity of the MAGIC stereo system is best  above 250 - 300\, GeV. For sources with spectral shapes softer than that of the Crab Nebula, the best performance occurs at slightly lower energies.
Consequently, we chose 250\,GeV as the minimum energy to report signal significances and $\gamma$-ray fluxes in light curves, while for the spectral analysis, in order to use all the available information, we also considered energies well below 250\,GeV, where the analysis of the MAGIC data can still be performed.

\subsection{\textit{Fermi}-LAT data analysis}

1ES 0806+524 has been observed by the pair conversion telescope \textit{Fermi}-LAT optimised for energies from 20\,MeV up to energies beyond 300\,GeV~\citep{atwood09}. In survey mode the \textit{Fermi}-LAT scans the entire sky every three hours.
The data sample, which consists of observations between 2010 November 22 and 2011 June 13, was analysed with the standard analysis tool {\it gtlike}, part of the \textit{Fermi} Science Tools software package (version 09-27-01) available from the \textit{Fermi} Science Support Center (FSSC)\footnote{\url{http://fermi.gsfc.nasa.gov/ssc/}}.

We selected P7CLEAN events located in a circular region of interest of $10^\circ$ radius centred on the position of 1ES 0806+524. To reduce the contamination from the Earth-limb $\gamma$ rays produced by cosmic-ray interactions with the upper atmosphere, data were restricted to a maximum zenith angle of $100^{\circ}$.

For the $\gamma$-ray signal extraction, the background model used included two components: Galactic diffuse and an isotropic diffuse emission, provided by the publicly available files gal\_2yearp7v6\_trim\_v0.fits and iso\_p7v6clean.txt\footnote{\href{http://fermi.gsfc.nasa.gov/ssc/data/access/lat/BackgroundModels.html}{http://fermi.gsfc.nasa.gov/ssc/data/access/lat/BackgroundModels.html}}. 

The model of the region of interest (ROI) also included sources from the second \textit{Fermi}-LAT catalog (2FGL;~\citealt{nolan12})  that are located within $15^\circ$ of 1ES 0806+524. These sources, as well as the source of interest, were modelled with a power law spectral shape with the initial parameters set to their 2FGL values. In the 2FGL catalog 1ES 0806+524 is associated with the source 2FGL J0809.8+5218, which has been reported with a flux of $(2.2\pm0.3)\times10^{-8}$\,cm$^{-2}$\,s$^{-1}$ and a photon index of  $(1.94\pm0.06)$ in the 2FGL catalog. When fitting, the spectral parameters of sources within $10^{\circ}$ from our target were allowed to vary while those within $10^{\circ}-15^{\circ}$ were fixed to their initial values. 

During the spectral fitting, the normalizations of the background models were allowed to vary freely.  Spectral parameters were estimated from 300\,MeV to 300\,GeV using an unbinned maximum likelihood technique~\citep{mattox96} taking into account the post-launch instrument response functions (IRF; specifically P7CLEAN\_V6,~\citealt{ackermann12}). 

During the MAGIC observing period, the source was not significantly detected on a daily basis. To ensure a good compromise between having a significant detection in most of the intervals
and details on the temporal behavior of the source, the light curve was produced with weekly binning. Flux upper limits at 95 per cent confidence level were calculated for each time bin where the test statistic (TS\footnote{The test statistic value quantifies the probability of having a pointlike $\gamma$-ray source at the location specified. It corresponds roughly to the standard deviation squared assuming one degree of freedom~\citep{mattox96}. The TS is defined as $-2\log (L_0 / L)$, where $L_0$ is the maximum likelihood value for a model without an additional source (i.e. the 'null hypothesis') and $L$ is the maximum likelihood value for a model with the additional source at the specified location.}) value for the source was below 9 (see section\,\ref{subsec:MWL}).

The systematic uncertainty in the flux is dominated by the systematic uncertainty in the effective area, which is estimated to be 10\% at 100\,MeV, decreasing to 5\% at 560\,MeV, and increasing to 10\% at 10\,GeV~\citep{ackermann12}. The systematic uncertainties are smaller than the statistical uncertainties of the data points in the light curve and spectra.

\subsection{\textit{Swift} observations and analysis}

Beside its prime objective of detecting and following up $\gamma$-ray bursts, since its launch in 2004 November the \textit{Swift} Gamma-Ray Burst observatory has become an instrument suitable for various purposes due to its fast response and its MWL abilities~\citep{gehrels04}. \textit{Swift} hosts three telescopes optimised for different energy ranges: the Burst Alert Telescope (BAT;~\citealt{barthelmy05}) suited for observations between 15-150\,keV, the X-ray telescope (XRT;~\citealt{burrows05}) with a 0.3-10\,keV coverage and the UV/Optical Telescope (UVOT;~\citealt{roming05}) optimal for observations within the 1800-6000\,$\mathring{\mathrm{A}}$ wavelength range.

Following the VHE $\gamma$-ray flare detection of 1ES 0806+524 by MAGIC~\citep{mariotti11}, \textit{Swift} ToO observations were requested and performed from 2011 February 26 to March 2; five observations together with MAGIC. A high-activity state of the source was confirmed reporting clear variability in X-rays~\citep{stamerra}. The source was monitored with the \textit{Swift}/XRT in photon counting (PC) mode with $\sim$2\,ks snapshots each night for a total exposure time of 10\,ks.

The data processing was performed with the  \texttt{xrtpipeline} v0.12.6 distributed by HEASARC as part of the HEASoft package. Events with grades 0-12 (according to the \textit{Swift} nomenclature;~\citealt{burrows05}) were selected for the PC data and the response matrices included in the \textit{Swift} CALDB\footnote{\url{http://heasarc.gsfc.nasa.gov/docs/heasarc/caldb/swift/}} were applied. All observations showed a source count rate $>$ 0.5 counts s$^{-1}$, thus pile-up correction was required. We extracted the source events from an annular region with an inner radius of 3 pixels (estimated by means of the point spread function (PSF) fitting technique, see~\citealt{moretti05}) and an outer radius of 30 pixels, while background events were extracted from an annular region centred on the source with radii of 70 and 120 pixels.  Ancillary response files were generated with \texttt{xrtmkarf}, and account for different extraction regions, vignetting and PSF corrections.

We fit the spectra in the 0.3-10\,keV energy range. The spectra were rebinned with a minimum of 20 counts per energy bin so that the $\chi^2$ minimization fitting technique could be used. The spectral analysis was performed with XSPEC~\citep{arnaud96} adopting a simple power law model that includes a hydrogen-equivalent column density fixed to the Galactic value $nH = 4.1\times10^{20}\,\mathrm{cm}^{-2}$~\citep{kalberla05}.

\textit{Swift}/UVOT data were taken with the `filter of the day' (either U, UVW1, UVM2, or UVW2 filter;~\citealt{poole08}) chosen day by day by the \textit{Swift} science planners. We analysed the data using the \texttt{uvotsource} task included in the HEASOFT package. Source counts were extracted from a circular region of 5\,arcsec radius centred on the source, while background counts were derived from a circular region of 10\,arcsec radius in the source neighborhood. Conversion of magnitudes into dereddened flux densities was obtained by adopting the extinction value $E(B-V) = 0.039$ from~\citet{schlegel98}, the mean Galactic extinction curve in~\citet{fitzpatrick99} and the magnitude-flux calibrations by~\citet{bessell98}.

\subsection{Optical observations and data analysis}

The optical data used in this study were obtained with the KVA\footnote{\url{http://www.astro.utu.fi/telescopes/60lapalma.htm}} located at the Roque de los Muchachos observatory on La Palma, and consisting of two telescopes mounted on the same fork. The smaller, a 35 cm Celestron, is used for photometric measurements, while the larger, a 60 cm one, is used for polarimetric observations. They are operated remotely from Finland. The photometric measurements are performed in the optical R-band using differential photometry, by having the target and the calibrated comparison stars in the same CCD images~\citep{fiorucci98}. The magnitudes of the source and comparison stars are measured using aperture photometry and converted to linear flux densities according to the formula $F(\mathrm{Jy}) = F_{0}\times10^{\mathrm{magR}/-2.5}$, where $F_{0}$ is a filter-dependent zero point ($F_{0}$ = 3080\,Jy in the R-band, from~\citealt{bessell79}). To obtain the AGN core emission, contributions from the host galaxy and possible nearby stars that add to the overall flux have to be subtracted from the obtained value. These values were determined by~\citet{nilsson07} and in the case of 1ES 0806+524 the host galaxy contribution of (0.69 $\pm$ 0.04)\,mJy was  subtracted from the overall flux.

The KVA telescope is operated under the Tuorla Blazar Monitoring Program\footnote{Project web page:~\url{http://users.utu.fi/kani/1m/}}, which has been running as a support program to the MAGIC observations since the end of 2002. The project uses the Tuorla 1\,m (located in Finland) and the KVA telescopes to monitor candidates (from~\citealt{costamante02}) and known TeV blazars in the optical waveband and to provide alerts to MAGIC on high states of these objects in order to trigger follow-up VHE observations. 1ES 0806+524 was one of the objects on the original target list and has therefore been monitored regularly since the beginning of the program. The object had been relatively dormant over the last years, showing some variability but no particular flaring activity. The large flare occurring at the end of 2010 (see section\,\ref{subsec:MWL}) is by far the largest optical activity recorded from this source during the last ten years, although this kind of flaring activity is not uncommon in the optical light curves of other blazars observed within the Tuorla monitoring program.

\subsection{OVRO data analysis}

Observations at 15\,GHz were carried out with the 40\,m OVRO radio telescope, located in California.
The observations of 1ES 0806+524 were carried out in the framework of a blazar monitoring program~\citep{richards11} measuring the source flux density twice a week. Occasional gaps in the data sample are due to poor weather conditions or maintenance. Observations were performed using a Dicke-switched dual-beam
system, with a second level of switching in azimuth to alternate between source and sky in each of the two horns, which removes much of the atmospheric and ground interference~\citep{readhead89}. The data were calibrated against 3C 286 with an assumed flux density of 3.44\,Jy at 15\,GHz~\citep{baars77}. The data were analysed using the pipeline described in~\citet{richards11}.

\section{Results}

\subsection{MAGIC results}

\begin{figure}
   \centering
\includegraphics[width=90mm]{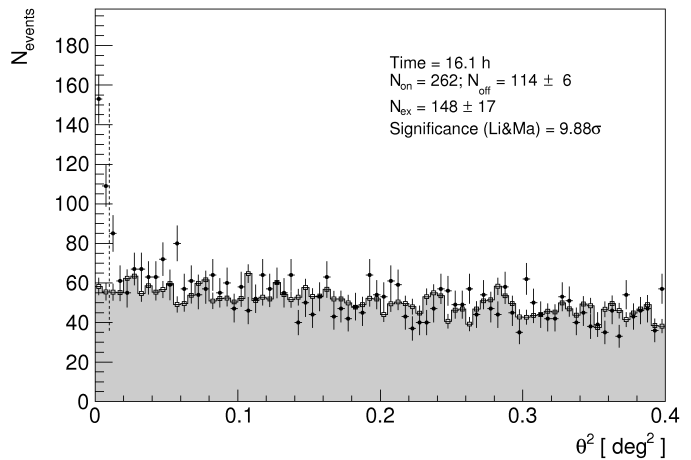}
   \caption{Distribution of the squared angular distance ($\theta^{2}$) for the on-source counts in the direction of 1ES 0806+524 (black points with error bars) and the normalised off-source events (gray histogram and open black squares) extracted from three background regions which are located at $90^\circ$, $180^\circ$ and $270^\circ$ with respect to the reconstructed source position in the camera. The signal is extracted in the $\theta^{2}$-region indicated by the vertical dashed line.}
    \label{fig:theta2}
    \end{figure}

Above an energy threshold of 250\,GeV the MAGIC data yield (after event selection cuts) an excess of $148\pm17$ events in the distribution of the squared angular distance $\theta^{2}$ between the reconstructed event direction and the catalog position of 1ES 0806+524, i.e. in the so-called `On' region. The background level of $114\pm6$ events was calculated from three equivalent `Off' regions, located at $90^\circ$, $180^\circ$ and $270^\circ$ with respect to the reconstructed source position in the camera, applying the same event selection cuts (Fig.~\ref{fig:theta2}).
    
The significance of the event excess corresponds to 9.9\,$\sigma$ calculated with formula [17] of Li \& Ma (1983). During the MAGIC observations, the source underwent a flaring event on February 24~\citep{mariotti11}. Within 3.0\,h of observation an excess of $50\pm 8$ events ($15\pm2$ background events) above 250\,GeV was measured corresponding to a confidence level of 7.6\,$\sigma$. After excluding the flare of February 24 from the data set, the remaining MAGIC observations (13.1\,h) still show a significant detection of $96\pm15$ excess events ($99\pm6$ background events) above 250\,GeV corresponding to 7.3\,$\sigma$. In the following, we will refer to the VHE $\gamma$-ray flare as the high state. The  low state refers to a low activity in VHE $\gamma$ rays including the remaining MAGIC observations.
   \begin{figure}
     \centering
 \includegraphics[width=86mm]{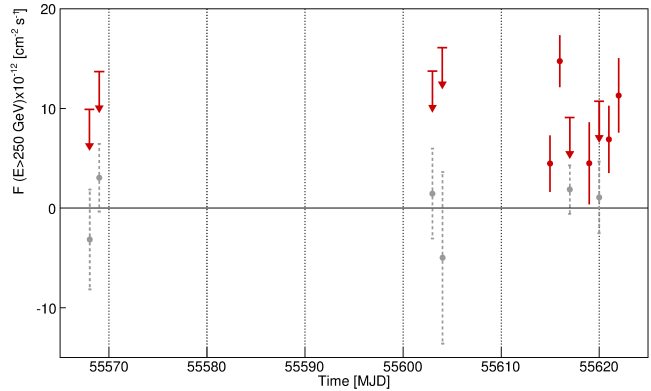}
   \caption{VHE $\gamma$-ray flux (E $>$ 250\,GeV) of 1ES 0806+524 for the single MAGIC observations performed in 2011. The red arrows correspond to 95 per cent confidence upper limits, which were computed for the observations where the interval flux $\pm$ error contains zero. The measured fluxes for these intervals are also shown  (gray circles).}
              \label{fig:LC_VHE}%
    \end{figure}
    
The integral flux above 250\,GeV was $(5.9\pm1.1)\times10^{-12}$\,cm$^{-2}$\,s$^{-1}$ corresponding to $(3.7\pm0.7)$ per cent C.U. Excluding the flare, the integral flux was measured to be $(3.1\pm1.0)\times10^{-12}$\,cm$^{-2}$\,s$^{-1}$ corresponding to $(1.9\pm0.7)$ per cent C.U. ($2.1\pm0.7$ per cent C.U. above 300\,GeV). This result is in good agreement with the flux level above 300\,GeV reported by VERITAS in 2009 (1.8 per cent C.U.;~\citealt{acciari09}). During the night when the source was flaring the integral flux was $(1.5\pm0.3)\times10^{-11}$\,cm$^{-2}$\,s$^{-1}$ equal to $(9.3\pm1.9)$ per cent C.U. The flare showed a flux of $\sim$3 $\sigma$ above the mean flux level\footnote{The deviation from the constant flux level was computed by dividing the difference of the flux measured during the VHE $\gamma$-ray flare and the mean flux level by the sum in quadrature of the statistical errors on the flux in the high and mean flux levels.} which corresponded to a flux increase of about a factor of three (Fig.\,\ref{fig:LC_VHE}). No intra-night variability was found during the VHE $\gamma$-ray flare within the statistical and systematic uncertainties. Besides the first flare observed by MAGIC on 2011 February 24 (MJD = 55616), the VHE data showed a hint of increasing flux towards March 2 (MJD = 55622), when the source was detected at $\sim$5$\,\sigma$ with a flux of $(7.0\pm2.3)$ per cent C.U. above 250\,GeV. Fitting the overall light curve with a constant function yields a probability of 0.5 per cent ($\chi^2$/d.o.f. = 24.92/10) for a non-variable source. Since the nights before and after the flaring event of February 24 showed a significantly lower flux, we assume short-term variability of the timescale of one day as an upper limit.
 \begin{figure}
   \centering
\includegraphics[width=90mm]{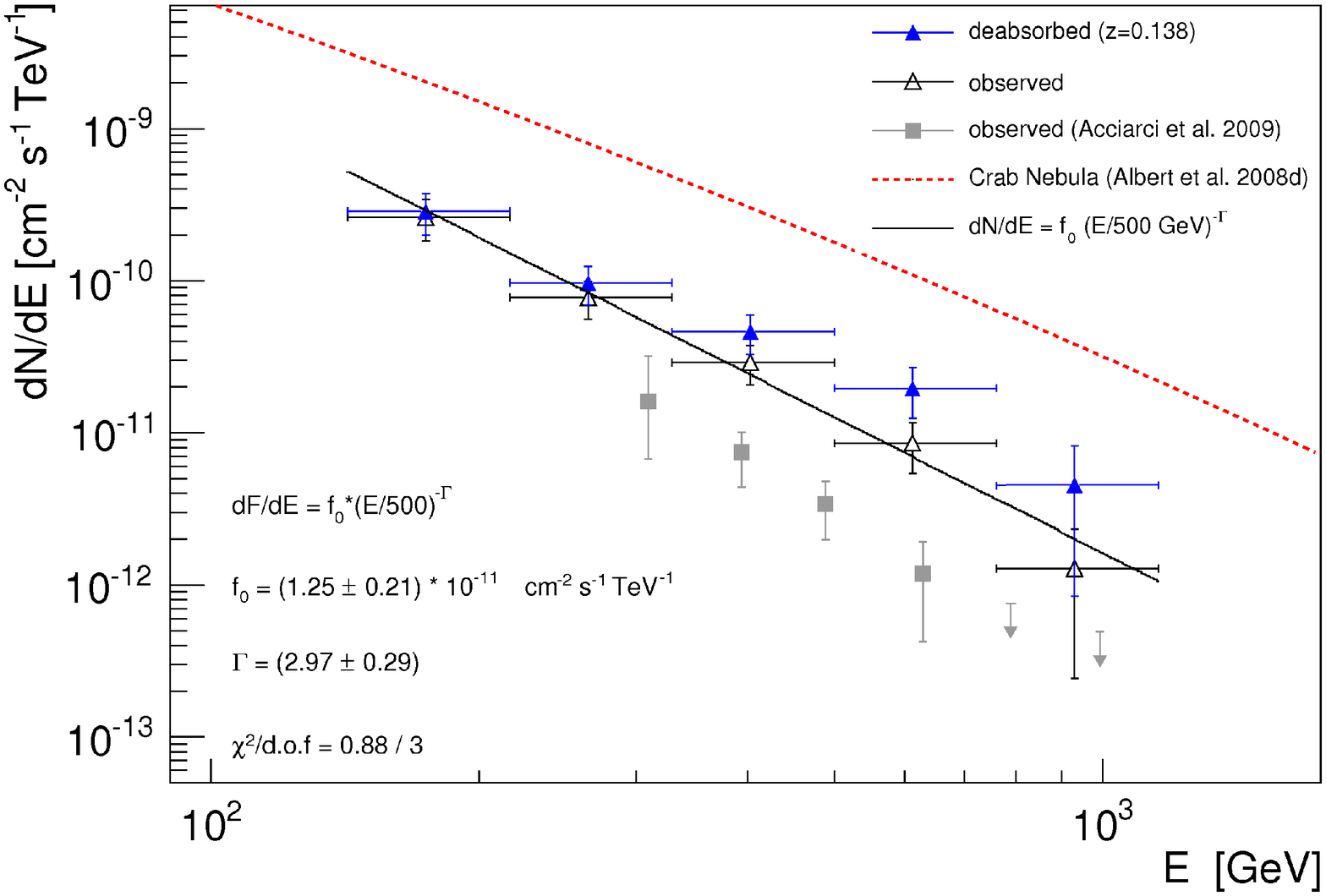}
   \caption{Unfolded high-state differential energy spectrum of 1ES 0806+524 measured by MAGIC on 2011 February 24. The open black triangles correspond to the measured spectrum to which a simple power law (solid black line) is fitted; the filled blue triangles depict the measured spectrum after correction for the EBL attenuation using the model by~\citet{dominguez11}. For comparison, the observed spectrum and derived upper limits (gray squares and arrows respectively) published by VERITAS ~\citep{acciari09} and the Crab Nebula spectrum (red dashed line) published by MAGIC ~\citep{albert08d} are shown. See text for further details.}
           \label{fig:spectrum1}
    \end{figure}

\begin{figure}
     \centering
\includegraphics[width=90mm]{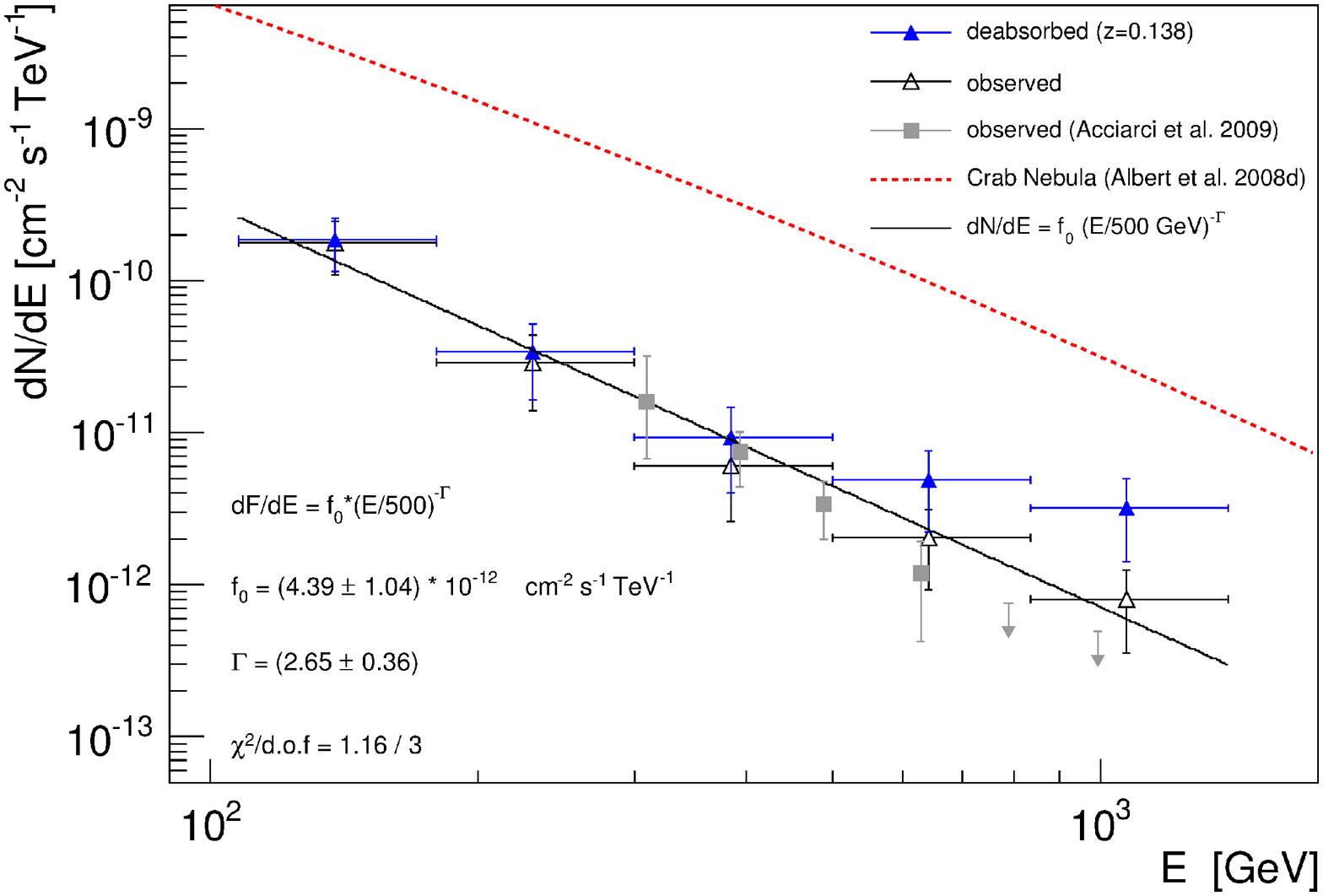}
   \caption{Unfolded low-state differential energy spectrum of 1ES 0806+524 measured by MAGIC in 2011. The open black triangles correspond to the measured spectrum to which a simple power law (solid black line) is fitted; the filled blue triangles depict the measured spectrum after correction for the EBL attenuation using the model by~\citet{dominguez11}. For comparison, the observed spectrum and derived upper limits (gray squares and arrows respectively) published by VERITAS ~\citep{acciari09} and the Crab Nebula spectrum (red dashed line) published by MAGIC ~\citep{albert08d} are shown.}
           \label{fig:spectrum2}
    \end{figure}

The differential energy spectra of the high and low states are shown in Figs.~\ref{fig:spectrum1} and~\ref{fig:spectrum2} respectively. Both spectra can be described with a simple power law\footnote{The fits of the high and low state spectra with a power law have a $\chi^2$/d.o.f. of 0.88/3 (83\,\%)  and 1.16/3 (76\,\%), respectively.} of the form $dN/dE=f_{0}(E/500\,\mathrm{GeV})^{-\Gamma}$ with a flux normalization $f_{0}$ at 500\,GeV of $(1.25\pm0.21)\times10^{-11}$\,cm$^{-2}$\,s$^{-1}$\,TeV$^{-1}$ for the high and $(4.39\pm1.04)\times10^{-12}$\,cm$^{-2}$\,s$^{-1}$\,TeV$^{-1}$ for the low state. The photon index $\Gamma$ was found to be $2.97\pm0.29$ for the high and $2.65\pm0.36$ for the low state. The spectra have been unfolded using the Tikhonov algorithm to correct for the finite energy resolution of MAGIC. Different unfolding algorithms as described in~\citet{albert07b} were compared and found to agree within errors.

To account for Extragalactic Background Light (EBL) attenuation, a correction by means of the EBL  model of~\citet{dominguez11}\footnote{This model is among the most recent ones (e.g.~\citealt{franceschini08,finke10,gilmore12}) and agrees within the uncertainties with other models.}
was applied. The corrected spectral flux of the two different source states follows a power law parametrized by a photon index $\Gamma=2.30\pm0.52$ and $2.15\pm0.59$ as well as a flux normalization $f_{0}$ at 500\,GeV $(2.57\pm0.81)\times10^{-11}$\,cm$^{-2}$\,s$^{-1}$\,TeV$^{-1}$ and $(9.10\pm4.03)\times10^{-12}$\,cm$^{-2}$\,s$^{-1}$\,TeV$^{-1}$, for the high (Fig.~\ref{fig:spectrum1}) and low state (Fig.~\ref{fig:spectrum2}), respectively. The low state spectrum observed by MAGIC shows a good agreement within errors with the observed spectral points published by VERITAS. A comparison of the differential energy spectra describing the different source states indicates an increase in flux but no significant hardening of the spectrum.

\subsection{Multiwavelength results}\label{subsec:MWL}

\begin{figure*}
\centering
\includegraphics[width=18.5cm]{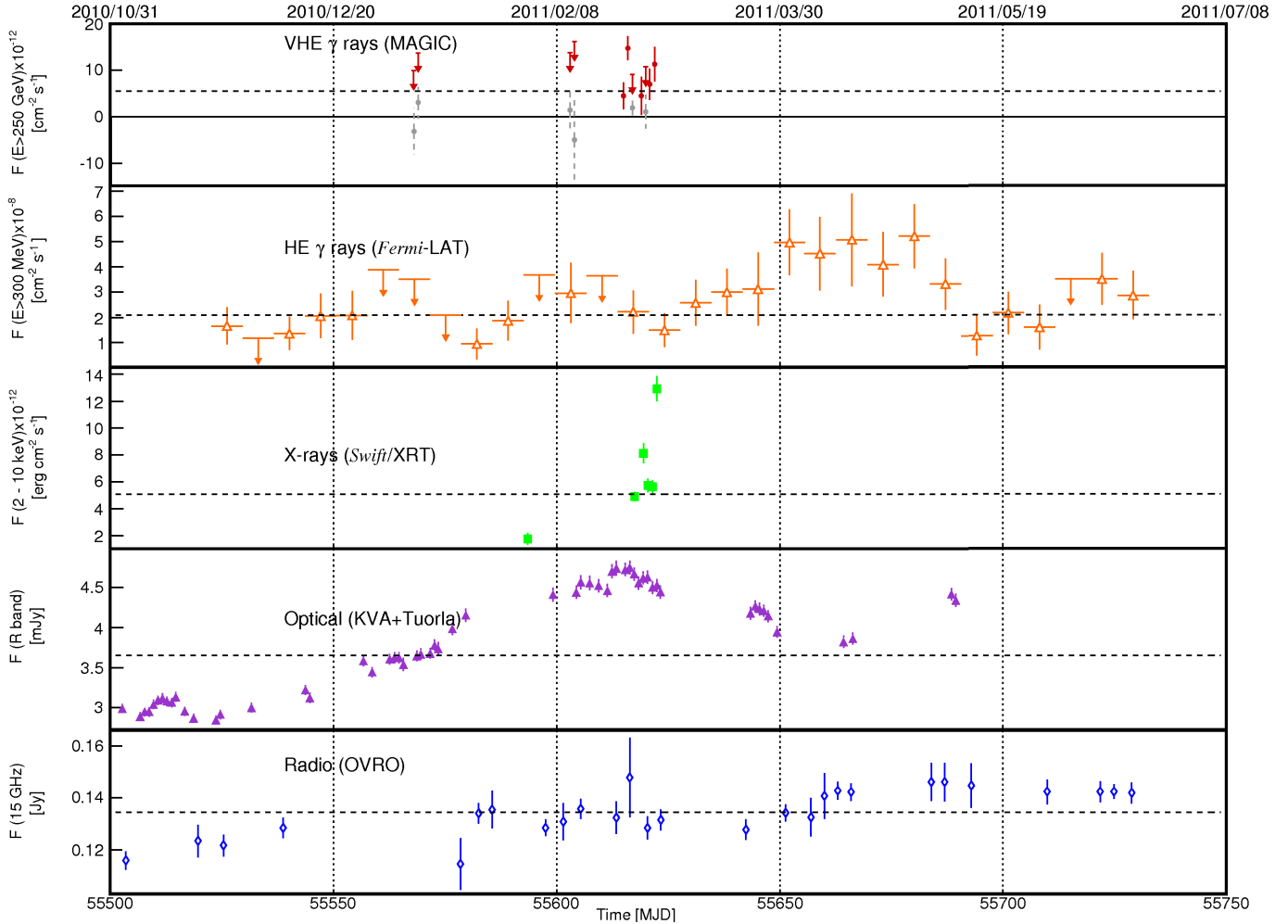}
\caption{MWL light curves of 1ES 0806+524 from 2010 December to 2011 June. From top to bottom: VHE $\gamma$ rays (red circles) by MAGIC, HE $\gamma$ rays (orange triangles) by \textit{Fermi}-LAT, X-rays (green squares) by Swift/XRT, R-band (purple triangles) by KVA, and radio (blue diamonds) by OVRO. The optical data are corrected for the host galaxy contribution according to Nilsson et al. (2007). Upper limits at 95 per cent confidence level are indicated by downward arrows. In the VHE light curve, upper limits (red arrows) are computed for the observations where the interval flux $\pm$ error contains zero. The measured fluxes for these intervals are also shown  (gray circles). The individual light curves are daily binned except for the light curve in the HE band from Fermi, where a binning of seven days has been used. The horizontal dashed lines report the mean flux in each light curve.}
\label{fig:LC_MWL}
\end{figure*}

Figure~\ref{fig:LC_MWL} shows the long-term MWL light curves in VHE $\gamma$ rays (MAGIC), HE $\gamma$ rays (\textit{Fermi}-LAT), X-rays (\textit{Swift}), in the R-band (KVA, Tuorla) and radio regime (OVRO telescope).

The  \textit{Fermi}-LAT light curve was produced fixing the power law index of the source of interest to the value derived from the spectral analysis of the data set from 2010 November 22 to 2011 June 13, i.e. $1.88\pm0.17$.

Possible variations in the source emission in HE $\gamma$ rays have been tested following the same likelihood method described in the 2FGL~\citep{nolan12}. The method, applied to the data from 2010 November to 2011 June, indicates that the source is variable (TS$_{var}=52$ for 29 d.o.f.). When applied only to data taken during the MAGIC observation period (MJD = 55603 to MJD = 55622) the probability for constant emission increases to 23 per cent (TS$_{var}$/d.o.f. = 10.5/8). Considering the entire data sample, i.e. from 2010 November 22 to 2011 June 13, the average flux level corresponds to $(2.10\pm 0.14)\times10^{-8}$\,cm$^{-2}$\,s$^{-1}$ (E $>$ 300\,MeV).

A smooth flux increase from the beginning of 2011 March until 2011 mid-April reaching a maximum of $(5.2\pm1.3)\times10^{-8}$\,cm$^{-2}$\,s$^{-1}$ was observed with a delay compared to the other wavelengths. Given the long integration time of seven days, no clear conclusion regarding simultaneous source variability with respect to that observed in VHE $\gamma$ rays can be drawn within the time interval of the flare. In particular, during the night of the VHE $\gamma$-ray flare the source was only marginally detected (TS = 4) in the \textit{Fermi}-LAT energy band.

In order to provide simultaneous coverage in this energy band, we computed the 95 per cent  confidence level upper limit between 300\,MeV and 300\,GeV for February 24 corresponding to $7.2\times10^{-11}$\,erg\,cm$^{-2}$\,s$^{-1}$ at a center energy of 580\,MeV.

The X-ray spectra derived with the individual \textit{Swift}/XRT observations are reported in Table~\ref{tab:XRT}. The time-averaged X-ray flux in the band 2-10\,keV is $(5.1\pm0.2)\times10^{-12}$\,erg\,cm$^{-2 }$\,s$^{-1}$,  with a flux increase by about a factor of six when comparing the lowest flux (occurring on February 1) with the highest flux (occurring on March 2). A fit with a constant function yields $\chi^2$/d.o.f. = 150/5, showing a clear X-ray variability and confirming the high-activity state of the source during the MAGIC observations performed at the end of February and beginning of March.

\begin{table}
\caption{Log and fitting results of {\em Swift}/XRT observations of 1ES 0806+524 using a power-law model with $N_{\rm H}$ fixed to Galactic absorption in the range from 0.3 to 10\,keV.}
\begin{tabular}{cccc}
\hline \hline
\multicolumn{1}{c}{Observation} &
\multicolumn{1}{c}{Photon index} &
\multicolumn{1}{c}{Flux 2-10 keV} &
\multicolumn{1}{c}{$\chi^2$/d.o.f.} \\
\multicolumn{1}{c}{Date [MJD]} &
\multicolumn{1}{c}{$\Gamma$}&
\multicolumn{1}{c}{$\times$10$^{-12}$ erg cm$^{-2}$ s$^{-1}$}\\
\hline
 55593  & $2.75  \pm 0.30$& $1.76 \pm 0.43$& Cash$^a$\\
   55617 &$2.36   \pm 0.09$ & $4.91 \pm 0.36$  &  36.84/34\\
   55619 & $2.44   \pm 0.10$& $8.13 \pm 0.73$ &  27.73/27\\
   55620 &  $2.38  \pm  0.09$& $5.73 \pm 0.51$&46.39/37\\
   55621 &   $2.40  \pm  0.11$& $5.63 \pm 0.49$&33.67/27\\
    55622 & $2.29   \pm 0.08$& $12.92 \pm 0.94$&38.58/40\\
\hline
\hline
\multicolumn{4}{l}{$^a$: The Cash statistic~\citep{humphrey09} was used to fit the}\\
\multicolumn{4}{l}{spectrum.}
\end{tabular}
\label{tab:XRT}
\end{table}

\textit{Swift}/UVOT observations were carried out with different filters in the ultraviolet bands. The brightness of (14.4 $\pm$ 0.03)\,mag measured on March 2 in the UVW2 and UVM2 bands is almost unchanged with respect to February 1 where a brightness of $(14.5\pm0.03)$\,mag was measured. However, 1ES 0806+524 appears about 1\,mag brighter compared to the UV flux observed in 2008 March. During the observations, the UV band photometry is compatible with a constant flux within the errors.

\begin{figure*}
\centering
   \includegraphics[width=18cm]{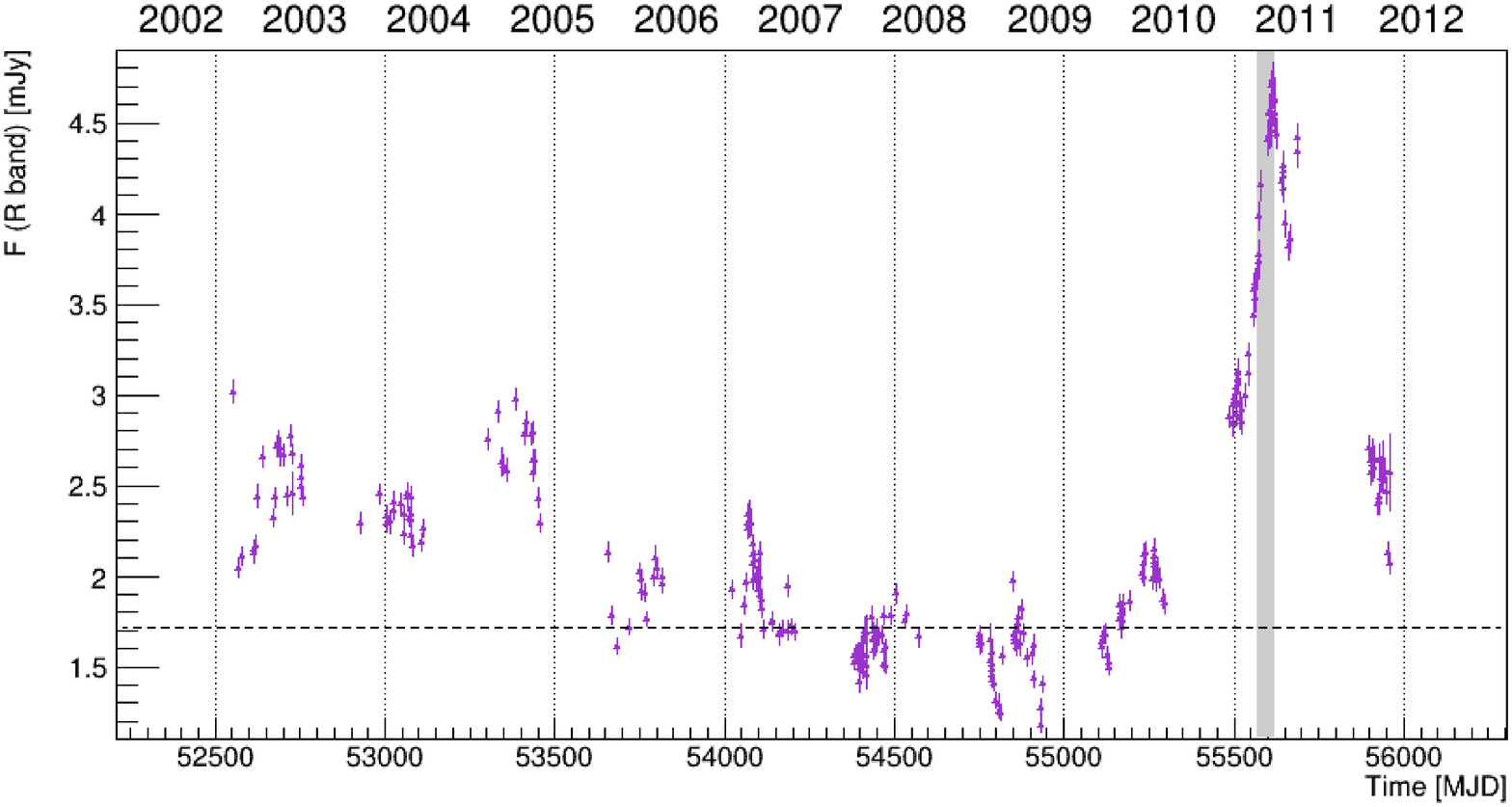}
   \caption{Host-galaxy-corrected~\citep{nilsson07} long-term light curve of 1ES 0806+524 of the optical R-band observed by the KVA telescope. The gray shaded area marks the period of the MAGIC observations in 2011. The quiescent state~\citep{reinthal12} is indicated as a dashed line.}
                \label{fig:LC_KVA}
     \end{figure*}

In the R-band, the core flux showed a large increase over the long-term base level starting from  2010 November (Fig.~\ref{fig:LC_KVA}). In the subsequent months the flux density continued to increase until reaching a maximum of $ (4.75 \pm 0.09)$\,mJy (host galaxy subtracted) in the night of 2011 February 24 during the outburst in VHE  $\gamma$ rays; this was almost three times higher than the quiescent state of 1.72\,mJy~\citep{reinthal12}. Later, the flux level steadily decreased.

The long-term radio light curve is consistent with constant emission (probability of $\sim$6.2$\times10^{-5}$; $\chi^2$/d.o.f. = 61.25/25) with an average flux level of $(0.136\pm0.01)$\,Jy. Considering only data taken during MAGIC observations (MJD = 55603 to MJD = 55622), the hypothesis of a non-variable emission at a mean flux level of 0.13\,Jy has a probability of 75 per cent. Compared to 2010 November observations, the radio data show a marginal flux increase from 2011 mid-January to May, exceeding the mean flux level of the overall observation period.

\subsection{Variability study across wavebands}
To study the connection between variability patterns in individual wavebands, MWL data gathered during the MAGIC observations (from MJD 55568 to MJD 55622) were studied. Due to the sparse overlap of simultaneous multi-frequency data and the limited duration of the MWL observations, detailed methods such as e.g. calculating cross-correlation functions of simultaneous data sets were not applicable. We therefore settled for a simple linear 2D-regression analysis where we plotted the simultaneous data points from one waveband as a function of flux in the comparison waveband. 

MWL data from five different wavebands (radio, optical, X-rays, HE $\gamma$ rays and VHE $\gamma$ rays) were available for comparison. 
The HE $\gamma$-ray band was excluded due to the inability to detect 1ES 0806+524 with \textit{Fermi}-LAT on day timescales, as it was done in the other wavebands. The data in the optical, X-rays and VHE $\gamma$ rays were fairly simultaneous being on average less than 2\,h and individually no more than 3.25\,h apart, while they all had an offset of roughly 12\,h with respect to the radio observations. In the case of two radio measurements (MJD = 55616.4 and MJD = 55620.4) there were two observations in other wavebands, performed with a gap of $\sim$12\,h before and after the corresponding radio observation. In those cases, we calculated the estimated flux level of the comparison waveband at the time of the radio measurement assuming linear behavior of the flux between the two points. This led to the exclusion of radio to X-ray and radio to VHE data sets from the study due to a low number of concurrent data points. In total, we have four waveband pairs for the study: radio vs. optical, optical vs. X-rays, optical vs. VHE $\gamma$ rays and X-rays vs. VHE $\gamma$ rays.

\begin{figure*}
   \centering
\includegraphics[width=18cm]{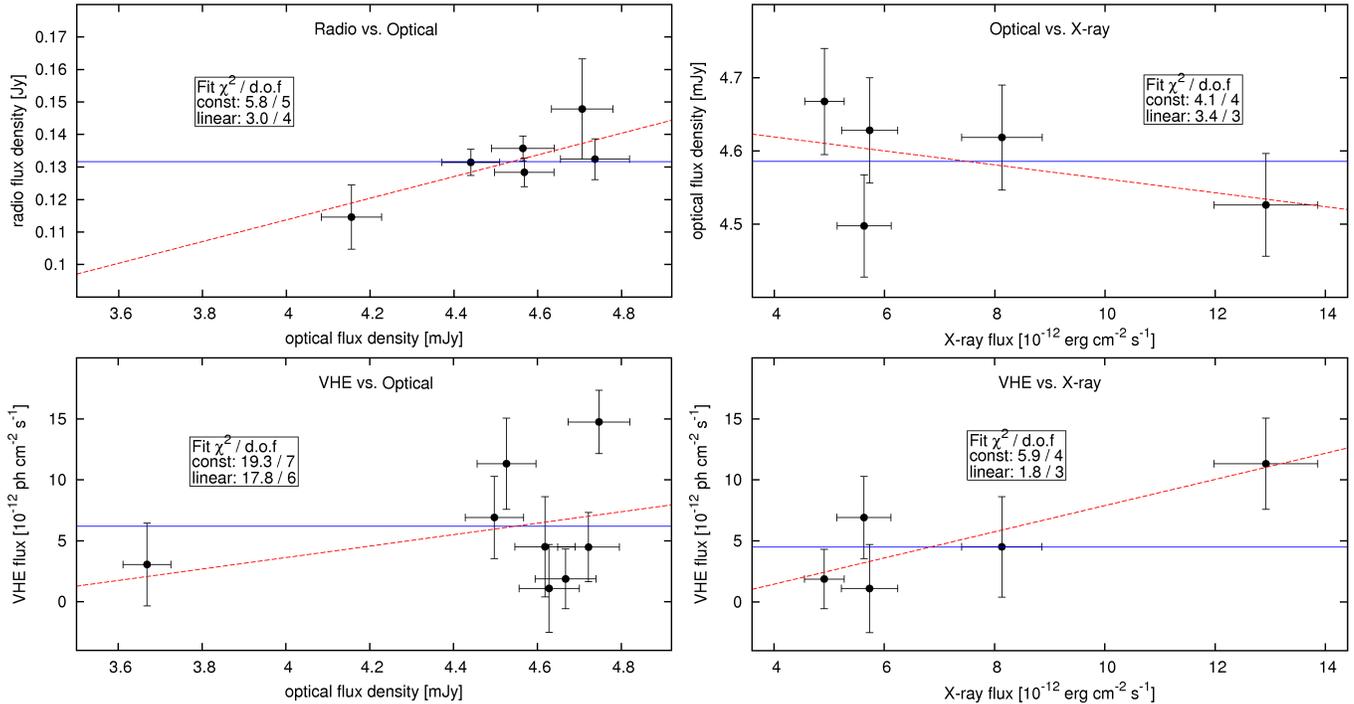}
\caption{One-to-one connection plots derived for the MWL correlation studies carried out between VHE $\gamma$ rays, X-rays as well as the optical and radio bands. The data are marked in black (filled circles), while the constant and the linear fits are represented by blue solid and red dashed lines respectively.}
\label{fig:mwl_corr}
\end{figure*}

We settled on a 16\,h window as a trade-off between strict simultaneity and feasible daily overlap for selecting the MWL correlation points. In practice, this only affected the comparison of radio to other wavebands as all the rest of them had fairly simultaneous pointings. Nevertheless the number of available points was quite low ranging from three to eight with radio to X-rays and radio to VHE having the lowest number of concurrent points. The result of this correlation study is shown in Fig~\ref{fig:mwl_corr}. 

In order to determine whether there is a correlation between activities in different wavebands we compared the probability of a linear fit to the data to that of a constant fit (Table~\ref{tab:regr_fit_prob}). 

\begin{table}
 \caption{Results of the linear regression analysis.}
 \label{tab:regr_fit_prob}
\begin{tabular}{l | c c c c}
\hline
\hline
 & optical to & optical to & optical to & X-ray to\\
 & radio & X-ray & VHE & VHE\\
\hline
   $\chi^2$ / d.o.f (const. fit) & 5.8 / 5 & 4.1 / 4 & 19.3 / 7 & 5.9 / 4 \\
   $\chi^2$ / d.o.f (lin. fit) & 3.0 / 4 & 3.4 / 3 & 17.8 / 6 & 1.8 / 3 \\
   fit likelihood (const. fit) & 0.33 & 0.39 & 0.0073 & 0.21 \\
   fit likelihood (lin. fit) & 0.56 & 0.34 & 0.0068 & 0.62 \\
   \hline
\hline
\end{tabular}
\end{table}

We find that the linear fits are not significantly better than the constant fits. The case with the largest difference occurs for the X-ray to VHE flux, where a likelihood ratio test shows that the constant fit has a tail probability (computed using the WilksÕ theorem) of 0.04.

It is worth noting that, while the VHE observations were triggered by the optical high state, there seems to be no one-to-one correlation between the fluxes in the two wavebands in the short term. Focussing on observations right before and after the flare, optical observations indicate a rather constant flux while the measurements performed in VHE $\gamma$ rays lead to the derivation of a short-term variability on a daily timescale. This suggests that the variability timescales are very different in these two bands and for future studies a longer time span should be used.

\section{Spectral energy distribution and interpretation}

\begin{figure*}
\centering
\includegraphics[width=18cm]{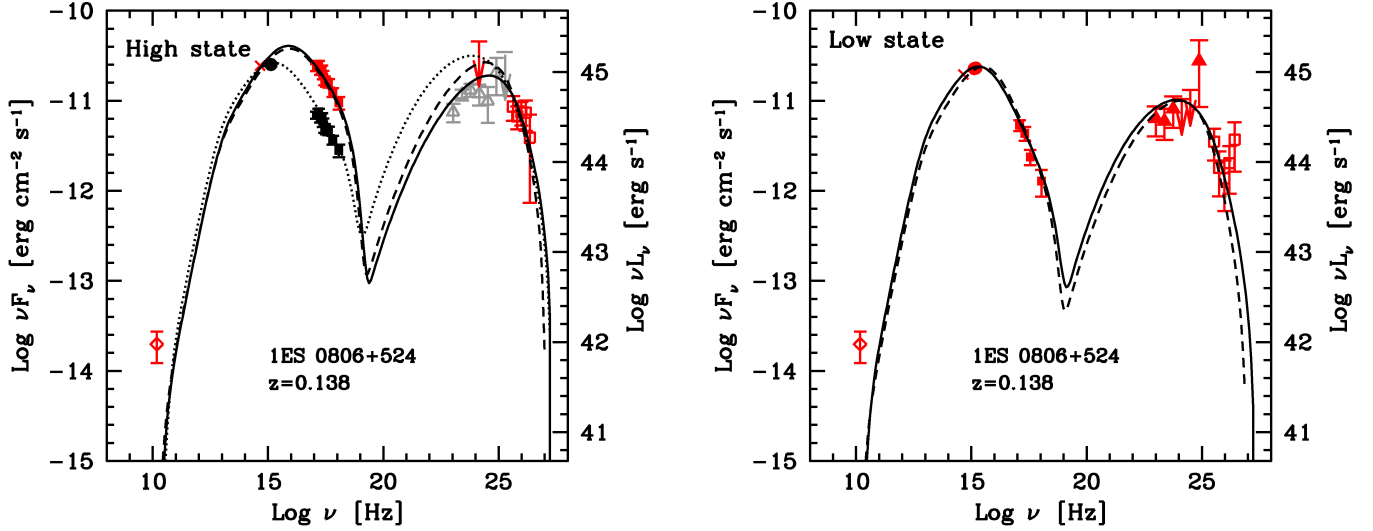}
\caption{SED of the high (left) and low (right) state of 1ES 0806+524 obtained during the MAGIC observations. The red (black) markers depict the simultaneous (quasi-simultaneous from February 25) MWL data. The MAGIC VHE $\gamma$-ray data are shown with open squares, \textit{Fermi}-LAT HE $\gamma$-ray data with triangles and arrows for the upper limits, \textit{Swift}-XRT data with filled squares, \textit{Swift}-UVOT data with filled circles, KVA data with crosses and OVRO data with open diamonds. The solid (dotted) black curve depicts the one-zone SSC model matching the simultaneous (quasi-simultaneous) data assuming a high Doppler factor. We also show the modelling (dashed line) of the simultaneous data assuming a lower Doppler factor. For comparison purposes, the left panel shows also the \textit{Fermi}-LAT HE $\gamma$-ray spectrum from 2010 November to 2011 June (grey open triangles and grey arrows). See text for further details.}
\label{fig:SED}
\end{figure*}

The SED of the source describing the high and low source state during the MAGIC observations, together with contemporaneous data from \textit{Fermi}, \textit{Swift}, the KVA and OVRO telescopes are presented in Fig.~\ref{fig:SED}. EBL corrections have been applied to the VHE $\gamma$-ray data using the model by~\citet{dominguez11}. \textit{Swift}/UVOT data and optical data in the R-band provided by the KVA telescope have been corrected for Galactic extinction and the host galaxy contribution respectively according to~\citet{fitzpatrick99} and~\citet{nilsson07}. 
Simultaneous and quasi-simultaneous data have been combined accordingly to the high and low state in VHE $\gamma$ rays. In general, quasi-simultaneous data have to be considered when studying the broad-band variability of blazars because strictly simultaneous observations are not always available. In this particular study, by quasi-simultaneous observations we mean the usage of \textit{Swift} (X-ray and UV) data from one day after the VHE flare detected by MAGIC.

Unfortunately, the high state data do not include a simultaneous HE $\gamma$-ray detection, as \textit{Fermi} did not significantly detect 1ES 0806+524 during the flares in VHE $\gamma$ rays and X-rays. In order to provide simultaneous coverage in this energy band, the 95 per cent  confidence level upper limit (300\,MeV to 300\,GeV) for 2011 February 24 was included in the SED modeling (due to the poor statistics only one upper limit for the whole energy band has been calculated). In addition, an averaged SED from eight months of observations taken from 2010 November to 2011 June has been derived and is shown for comparison. Consequently, the latter data are not included in the modeling of the high state SED. For the low state of 1ES 0806+524, an averaged \textit{Fermi} spectrum was produced considering data collected during the period from 2011 January to March 2 and excluding the night of the VHE $\gamma$-ray flare (February 24) from the data sample.

With respect to the \textit{Swift} data, the high state comprises observations carried out on February 25 since it is the closest X-ray spectrum to the VHE flare detected by MAGIC on February 24, whereas the low state encompasses observations performed on February 1. Because of the high variability in the X-ray and VHE bands (especially during the flaring state), the lack of simultaneity between these two energy bands places uncertainty in the interpretation of the broad band SEDs within the one-zone SSC model. Thus, we estimate the X-ray spectrum for February 24 by scaling the spectrum from February 25 (Fig.~\ref{fig:LC_MWL}) with a factor of 3.3, which is derived from the linear fit to the X-ray-to-TeV-flux (Fig.~\ref{fig:mwl_corr}). Since we were not able to reproduce both the optical and VHE $\gamma$-ray SED points from previous MWL observations~\citep{acciari09}, they are not shown.

The MAGIC observations show a clear flux variability of about a factor of three in VHE $\gamma$ rays when comparing the high and low states (see Figs.~\ref{fig:spectrum1} and~\ref{fig:spectrum2}). 
The \textit{Fermi}-LAT light curve study indicated variability for the time period considered in this work (MJD = 55603 to MJD = 55622).

The Swift/XRT spectrum from February 25 is marginally harder than that from February 1, and the X-ray flux above 2\,keV from February 25 is about three times higher than that from February 1 (Table~\ref{tab:XRT}). The ratio between the assumed X-ray spectrum for February 24 (high state) and the flux of February 1 (low state) is about nine.

The flux in the R-band started to increase well before the flaring activity in the VHE regime and in X-rays (see Fig.\,\ref{fig:LC_MWL}). It reached a maximum level at an almost constant flux with only marginal variations towards the end of February when the VHE $\gamma$-ray flare occurred. Therefore, no clear variation between the high and low state SEDs is seen. While the inverse Compton peak indicates flux variability between both activity states, the synchrotron component shows only minor variation. Unfortunately, the weak detection in the HE $\gamma$-ray band limits the determination of the inverse Compton peak.

A one-zone SSC model is applied to reproduce the SEDs of both source states (for a detailed description of the model see~\citealt{maraschi03}), where a spherical emission region of radius $R$ is assumed, filled with a tangled magnetic field of intensity $B$. A population of relativistic electrons is approximated by a smoothed broken power law that is parametrized by the minimum  $\gamma_{\rm{min}}$, break $\gamma_{\rm{b}}$ and maximum Lorentz factor $\gamma_{\rm{max}}$, as well as by the slopes $n_1$ and $n_2$ before and after the break respectively. Relativistic effects are taken into account by the Doppler factor $\delta$. The emission is self-absorbed at radio frequencies. This implies that the jet decelerates from the source region to the outer regions where the bulk of the radio emission originates. Therefore, radio data are not included in the SED modelling.

It is well known (\citealt{tavecchio98}) that the one-zone SSC model is constrained once the basic SED parameters and the variability timescale $t_{\rm var}$ - related to the source radius - are known. In the case studied here only an upper limit on the variability timescale is derived, $t_{\rm var}\lesssim 1$ day. This timescale is much larger than the variability timescale that has been measured during some flaring episodes in BL Lac objects, especially in the TeV band, for which sub-hour variability has been observed \citep{gaidos96,albert07c,aharonian07}. This uncertainty on $t_{\rm var}$ prevents us from strongly constraining the model parameters. In particular, the Doppler factor $\delta$ and the magnetic field $B$. We thus provide two possible realizations of the model, corresponding to two different values of the Doppler factor ($\delta=30$ and $\delta=15$) bracketing the typical range of values found in TeV BL Lac modeling (e.g.~\citealt{tavecchio10}).

The physical parameters derived reproducing the SEDs of both source states are reported in Table~\ref{tab:SED}. To model the high state, we consider the quasi-simultaneous X-ray data from February 25 observations ($\delta=30$) and the assumed simultaneous X-ray spectrum ($\delta=30$ and $\delta=15$). The values of the parameters are similar to those inferred for other HBL objects (see e.g.~\citealt{tavecchio10}).

The differences between the high and the low state is mainly driven by the value of the Lorentz factor of the electrons dominating the emission at the synchrotron peak, $\gamma_{\rm b}$, which almost doubles from the low to the high state, while the other parameters have almost the same value (except for the electron normalization $K$ that, for the case $\delta=30$, halves from the low to the high state). As expected~\citep{tavecchio98} the magnetic field intensity is inversely proportional to the Doppler factor, being therefore larger in the case $\delta=15$. B field values of $0.05-0.1$\,G are commonly derived for HBLs (e.g.~\citealt{tavecchio10}).

We also report the power carried by the jet through electrons ($P_{\rm e}$), magnetic field ($P_{B}$) and protons ($P_{\rm p}$, derived assuming the presence of one cold proton per emitting electron). We note that while in FSRQ this choice is dictated by energetic requirements, in the case of BL Lac there are no stringent constraints on the composition (e.g.~\citealt{celotti08}). However, this choice has a rather small impact on the derived total power, which is dominated by the leptonic component for both values of $\delta$. The magnetic field and protons appear to contribute less to the jet power, which is mostly carried by the electrons. Its total value $P_{\rm jet}=P_{\rm e}+P_{B}+P_{\rm p}\sim$10$^{44}$ erg s$^{-1}$ is also typical (e.g.~\citealt{ghisellini10}). While the jet power carried by the electrons and protons doubles during the source flare, the magnetic field strength is reduced.

The Doppler factors assumed for the SED modelling reported in this paper ($\delta=15$ and $\delta=30$) are larger than those typically derived from radio observations by about one order of magnitude. This discrepancy is commonly found in the TeV BL Lac population~\citep{blasi13}. Such so-called Doppler crisis implies that the jet decelerates from the source region to the outer regions where the bulk of the radio emission originates (e.g.~\citealt{georganopoulos03,ghisellini05}).

The assumed one-zone model, although applied here to the average SED of the low state, is strictly valid only for snapshots, since the traveling and expanding blob is expected to change the emission properties because of adiabatic and radiative losses. As noted e.g. in~\citet{tagliaferri08}, the emission region may be a standing shock through which the jet plasma continuously passes.

\begin{table*}
\caption{Input model parameters for the high and low state SEDs shown in Fig.~\ref{fig:SED}. We report the minimum $\gamma_{\rm{min}}$, break $\gamma_{\rm{b}}$ and maximum Lorentz factor $\gamma_{\rm{max}}$ and the low- and high-energy slope of the electron energy distribution, the magnetic field intensity, the electron density, the radius of the emitting region and its Doppler factor. We also report the derived power carried by electrons, magnetic field, and protons (assuming one cold proton per emitting relativistic electron and the frequencies of the synchrotron and inverse Compton peak). (I):  high state with quasi-simultaneous X-ray data; (IIa/b): high state with assumed simultaneous X-ray data modelled with a high/low Doppler factor; (IIIa/b): low state modelled with a high/low Doppler factor.}
\label{tab:SED}
    \begin{tabular}{ccccccccccccccl}
      \hline
      \hline
      $\gamma _{\rm min}$ & $\gamma _{\rm b}$ & $\gamma _{\rm max}$ & $n_1$ & $n_2$ &$B$ & $K$ &$R$ & $\delta $ &$P_{\rm e}$ & $P_{B}$ & $P_{\rm p}$&$\nu_\mathrm{sync}$& $\nu_\mathrm{IC}$&\\
      $ [10^3]$ & [$ 10^4$] &[$ 10^5$]  &  & &[G] & [$ 10^3$ cm$^{-3}]$  & $[10^{16}$ cm] & &  [$10^{43}$ erg s$^{-1}$] & [$10^{43}$ erg s$^{-1}$]& [$10^{43}$ erg s$^{-1}$] &[Hz] &[Hz] &\\
      \hline
      $1$ & $1.65$ & $7$ & $2$ & $3.85$ & $0.05$ & $19.0$ & $1.12$ & $30$  & 79.9 & 0.1& 28.9&$10^{15.26}$ &$10^{23.84}$&(I)\\

$1$ & $3.2$ & $7$ & $2$ & $3.85$ & $0.06$ & \,\,\,$3.0$ & $1.70$ & $30$  & 34.3 & 0.35&\,\,\,11&$10^{15.97}$ &$10^{24.61}$&(IIa)\\

$1$ & $3.6$ & $7$ & $2$ & $3.85$ & $0.10$ & $\,\,\,2.6$ & $3.00$ & $15$  & 24.0 & 0.76& \,\,\,7.3&$10^{15.90}$ &$10^{24.48}$&(IIb)\\

 $1$ & $2.1$ & $7$ & $2$ & $4.15$ & $0.05$ & $\,\,\,4.0$ & $1.70$ & $30$  & 43.0 & 0.2& 14.3&$10^{15.33}$ &$10^{23.87}$&(IIIa)\\
      
        $1$ & $2.8$ & $6$ & $2$ & $4.30$ & $0.10$ & $\,\,\,1.8$ & $3.20$ & $15$  & 18.6& 0.9& \,\,\,5.8&$10^{15.56}$ &$10^{24.09}$&(IIIb)\\
      \hline
      \hline
    \end{tabular}
    
  \end{table*}

Assuming that the whole emission region $R_\mathrm{em}$  is causally connected on the observed timescale, the physical parameters derive a minimal variability timescale of $t_{\rm{var,min}}= [R_{\mathrm{em}}\cdot(1+z)](c\cdot\delta)\sim 0.3\,(0.9)$ days for the high state assuming either the high (low) Doppler factor. The values found are both perfectly compatible with the variability timescale of one day inferred from the VHE light curve.

The SSC model of the quasi-simultaneous data (Fig~\ref{fig:SED}) implies an increase of the inverse Compton peak by a factor of three for the high state SED with respect to the low state. When using the assumed simultaneous X-ray spectrum, such increase is excluded. Considering the assumed simultaneous X-ray data, the SSC model indicates a shift of the synchrotron peak and the inverse Compton peak towards higher frequencies and a synchrotron dominance rather than an equal peak flux of both energy bumps. Such synchrotron dominance is more pronounced when using the high Doppler factor for the modelling.

In comparison to the VHE data from~\citet{acciari09}, the inverse Compton peak is more constrained by the MAGIC data of the high source state. The comparison between the physical parameters obtained from the low state SED observed by MAGIC to MWL observations of 1ES 0806+524 in 2008 March 8 is not straightforward, as~\citet{acciari09} applied an SSC model whose electron spectrum is approximated by an unbroken power law. The synchrotron and inverse Compton peak positions (\,$\nu_\mathrm{sync}\approx 10^{16}$\,Hz, $\nu_\mathrm{IC}\approx 10^{24}$\,Hz, corresponding to $\sim$41\,eV and $\sim$13\,GeV) derived for the  SED from 2008~\citep{acciari09} and MAGIC observations presented in this paper are located at the same order of frequency.

In contrast to ~\citet{acciari09}, our SSC model implies a clear deviation from equipartition (with electrons dominating the total energy density), which is even more pronounced during the high state of the source. The origin of this difference with the VERITAS results is likely due to the different synchrotron peak frequency that, in the case of VERITAS, allowed a larger value of the magnetic field.

\section{Summary \& Discussion}
In this article we report on the MAGIC observations of the HBL object 1ES 0806+524 from 2010 December to 2011 June, which were triggered by an optical high state. MAGIC detected a one-day long flare with a VHE flux three times larger than the flux during the low state. This flaring episode lasted less than one day. From this relatively short, high-activity state, whose observation in weak sources like 1ES 0806+524 is rather rare, a short-term variability of one-day timescale has been inferred. This in turn sets constraints on the size of the emitting region during the high VHE state. Excluding the flare night, the VHE $\gamma$-ray flux from the source was in good agreement with the one observed by VERITAS in 2008, when the source was in a lower optical state. We present detailed spectra in the energy range from a few hundred\,GeV to one TeV, describing both the high and the low VHE state.

We studied the variability patterns of 1ES 0806+524 during the MAGIC VHE $\gamma$-ray observations across the available wavebands ranging from radio to VHE $\gamma$ rays in order to investigate the possible connection between the emission at different frequencies. The 2D linear regression we carried out did not reveal a connection between the flux levels in any of the possible waveband combinations. Although the VHE observations were optically triggered, no apparent evidence of a short-term correlation was found between these wavebands.

As the multifrequency data cover both the high and the quiescent VHE state activities of 1ES 0806+524, we performed one-zone SSC modelling of the SEDs obtained from the respective data sets assuming two different Doppler factors due to the uncertainty on the variability timescale. This model could adequately explain the broad-band emission during both source states using physical parameters similar to those from other HBL objects. Non-simultaneity of some of the MWL data hampered detailed studies on the broad-band variability of 1ES 0806+524.

\section*{Acknowledgements}

We would like to thank
the Instituto de Astrof\'{\i}sica de Canarias
for the excellent working conditions
at the Observatorio del Roque de los Muchachos in La Palma.
The support of the German BMBF and MPG,
the Italian INFN,
the Swiss National Fund SNF,
and the ERDF funds under the Spanish MINECO
is gratefully acknowledged.
This work was also supported
by the CPAN CSD2007-00042 and MultiDark CSD2009-00064 projects of the Spanish Consolider-Ingenio 2010 programme,
by grant 268740 of the Academy of Finland,
by the Croatian Science Foundation (HrZZ) Project 09/176 and the University of Rijeka Project 13.12.1.3.02,
by the DFG Collaborative Research Centers SFB823/C4 and SFB876/C3,
and by the Polish MNiSzW grant 745/N-HESS-MAGIC/2010/0.\\
The \textit{Fermi} LAT Collaboration acknowledges generous ongoing support
from a number of agencies and institutes that have supported both the
development and the operation of the LAT as well as scientific data analysis.
These include the National Aeronautics and Space Administration and the
Department of Energy in the United States, the Commissariat \`a l'Energie Atomique
and the Centre National de la Recherche Scientifique / Institut National de Physique
Nucl\'eaire et de Physique des Particules in France, the Agenzia Spaziale Italiana
and the Istituto Nazionale di Fisica Nucleare in Italy, the Ministry of Education,
Culture, Sports, Science and Technology (MEXT), High Energy Accelerator Research
Organization (KEK) and Japan Aerospace Exploration Agency (JAXA) in Japan, and
the K.~A.~Wallenberg Foundation, the Swedish Research Council and the
Swedish National Space Board in Sweden.

Additional support for science analysis during the operations phase is gratefully
acknowledged from the Istituto Nazionale di Astrofisica in Italy and the Centre National d'\'Etudes Spatiales in France.\\
The OVRO 40-m monitoring program is
supported in part by NASA grants NNX08AW31G
and NNX11A043G, and NSF grants AST-0808050
and AST-1109911.


 \vspace{0.5cm}
$^{1}$ {IFAE, Campus UAB, E-08193 Bellaterra, Spain}\\
$^{2}$ {Universit\`a di Udine, and INFN Trieste, I-33100 Udine, Italy}\\
$^{3}$ {INAF National Institute for Astrophysics, I-00136 Rome, Italy}\\
$^{4}$ {Universit\`a  di Siena, and INFN Pisa, I-53100 Siena, Italy}\\
$^{5}$ {Croatian MAGIC Consortium, Rudjer Boskovic Institute, University of Rijeka and University of Split, HR-10000 Zagreb, Croatia}\\
$^{6}$ {Max-Planck-Institut f\"ur Physik, D-80805 M\"unchen, Germany}\\
$^{7}$ {Universidad Complutense, E-28040 Madrid, Spain}\\
$^{8}$ {Inst. de Astrof\'isica de Canarias, E-38200 La Laguna, Tenerife, Spain}\\
$^{9}$ {University of \L\'od\'z, PL-90236 Lodz, Poland}\\
$^{10}$ {Deutsches Elektronen-Synchrotron (DESY), D-15738 Zeuthen, Germany}\\
$^{11}$ {ETH Zurich, CH-8093 Zurich, Switzerland}\\
$^{12}$ {Universit\"at W\"urzburg, D-97074 W\"urzburg, Germany}\\
$^{13}$ {Centro de Investigaciones Energ\'eticas, Medioambientales y Tecnol\'ogicas, E-28040 Madrid, Spain}\\
$^{14}$ {Institute of Space Sciences, E-08193 Barcelona, Spain}\\
$^{15}$ {Universit\`a di Padova and INFN, I-35131 Padova, Italy}\\
$^{16}$ {Technische Universit\"at Dortmund, D-44221 Dortmund, Germany}\\
$^{17}$ {Unitat de F\'isica de les Radiacions, Departament de F\'isica, and CERES-IEEC, Universitat}\\
$^{18}$ {Universitat de Barcelona, ICC, IEEC-UB, E-08028 Barcelona, Spain}\\
$^{19}$ {Japanese MAGIC Consortium, KEK, Department of Physics and Hakubi Center, Kyoto University, Tokai University, The University of Tokushima, ICRR, The University of Tokyo, Japan}\\
$^{20}$ {Finnish MAGIC Consortium, Tuorla Observatory, University of Turku and Department of Physics, University of Oulu, Finland}\\
$^{21}$ {Inst. for Nucl. Research and Nucl. Energy, BG-1784 Sofia, Bulgaria}\\
$^{22}$ {Universit\`a di Pisa, and INFN Pisa, I-56126 Pisa, Italy}\\
$^{23}$ {ICREA and Institute of Space Sciences, E-08193 Barcelona, Spain}\\
$^{24}$ {Universit\`a dell'Insubria and INFN Milano Bicocca, Como, I-22100 Como, Italy}\\
$^{25}$ {now at NASA Goddard Space Flight Center, Greenbelt, MD 20771, USA and Department of Physics and Department of Astronomy, University of Maryland, College Park, MD 20742, USA}\\
$^{26}$ {now at Ecole polytechnique f\'ed\'erale de Lausanne (EPFL), Lausanne, Switzerland}\\
$^{27}$ {now at Institut f\"ur Astro- und Teilchenphysik, Leopold-Franzens- Universit\"at Innsbruck, A-6020 Innsbruck, Austria}\\
$^{28}$ {now at Finnish Centre for Astronomy with ESO (FINCA), Turku, Finland}\\
$^{29}$ {also at INAF-Trieste}\\
$^{30}$ {also at ISDC - Science Data Center for Astrophysics, 1290, Versoix (Geneva)}\\
$^{31}$ {Department of Physics and Astronomy and the Bartol Research Institute, University of Delaware, Newark, DE 19716, USA}\\ 
$^{32}$ {INAF-IRA Bologna, Via Gobetti 101, I-40129, Bologna, Italy}\\
$^{33}$ {Agenzia Spaziale Italiana (ASI) Science Data Center, I-00044 Frascati (Roma), Italy}\\
$^{34}$ {Istituto Nazionale di Astrofisica - Osservatorio Astronomico di Roma, I-00040 Monte Porzio Catone (Roma), Italy}\\
$^{35}$ {Aalto University Mets\"ahovi Radio Observatory, Mets\"ahovintie 114, FI-02540 Kylm\"al\"a, Finland}\\
$^{36}$ {Cahill Center for Astronomy and Astrophysics, California Institute of Technology, Pasadena CA, 91125, USA}\\
$^{37}$ {National Radio Astronomy Observatory, P.O. Box 0, Socorro, NM 87801, USA}\\
$^{38}$ {Department of Physics, Purdue University, 525 Northwestern Ave, West Lafayette, IN 47907, USA}\\
\label{lastpage}
\end{document}